\pdfoutput=1
\documentclass[traditabstract]{aa} % for the abstract without structuration 
\usepackage{natbib}
\usepackage{graphicx}
\usepackage{txfonts}
\usepackage{supertabular}

\bibpunct{(}{)}{;}{a}{}{,} % to follow the A&A style
\begin{document}
\title{$Herschel$ deep far-infrared counts through Abell 2218 cluster-lens} 
\titlerunning{$Herschel$ deep far-infrared counts through Abell 2218 cluster-lens}
\authorrunning{Altieri et al.}

\author{B.\,Altieri \inst{1}, 
S.\,Berta \inst{2}, 
D.\,Lutz \inst{2}, 
J.-P.\,Kneib \inst{3}, 
L.\,Metcalfe \inst{1}, 
P.\,Andreani \inst{5,12},
H.\,Aussel \inst{4},
A.\,Bongiovanni \inst{6}, 
A.\,Cava \inst{6}, 
J.\,Cepa \inst{6}, 
L.\,Ciesla \inst{3},
A.\,Cimatti \inst{7}, 
E.\,Daddi \inst{4}, 
H.\,Dominguez \inst{9}, 
D.\,Elbaz \inst{4}, 
N.M.\,F\"orster Schreiber \inst{2}, 
R.\,Genzel  \inst{2}, 
C.\,Gruppioni \inst{9}, 
B.\,Magnelli  \inst{2}, 
G.\,Magdis  \inst{4}, 
R.\,Maiolino \inst{8}, 
R.\,Nordon  \inst{2}, 
A.M.\,P\'erez Garc\'ia \inst{6},  
A.\,Poglitsch  \inst{2}, 
P.\,Popesso  \inst{2}, 
F.\,Pozzi \inst{7}, 
J.\,Richard \inst{10},
L.\,Riguccini  \inst{4}, 
G.\,Rodighiero \inst{11}, 
A.\,Saintonge  \inst{2}, 
P.\,Santini \inst{8}, 
M.\,Sanchez-Portal\inst{1}, 
L.\,Shao  \inst{2}, 
E.\,Sturm  \inst{2}, 
L.J.\,Tacconi  \inst{2}, 
I.\,Valtchanov\inst{1}, 
M.\,Wetzstein  \inst{2}, 
and E.\,Wieprecht  \inst{2}
\thanks{$Herschel$ is an ESA space observatory with science instruments provided by 
European-led Principal Investigator consortia and with important 
participation from NASA.}
}	

\offprints{Bruno Altieri: bruno.altieri@sciops.esa.int}

\institute{
	Herschel Science Centre, European Space Astronomy Centre, ESA, Villanueva de la Ca\~nada, 28691 Madrid, Spain.
	\and
	Max-Planck-Institut f\"{u}r Extraterrestrische Physik (MPE), Postfach 1312, 85741 Garching, Germany.
	\and
	Laboratoire d'Astrophysique de Marseille, CNRS-Universit\'e de Provence, 38 rue F. Joliot-Curie, 13013 Marseille, France.
        \and
        Laboratoire AIM, CEA/DSM-CNRS-Univ Paris-Diderot, IRFU/SAp,
        B\^at.709, CEA-Saclay, 91191 Gif-sur-Yvette, France.
        \and
        ESO, Karl-Schwarzschild-Str. 2, D-85748 Garching, Germany.
        \and
        Instituto de Astrof{\'i}sica de Canarias \& Departamento de Astrof{\'i}sica, Universidad de La Laguna, Spain.
        \and
        Dipartimento di Astronomia, Universit{\`a} di Bologna, via Ranzani 1, 40127 Bologna, Italy.
        \and
        INAF - Osservatorio Astronomico di Roma, via di Frascati 33, 00040 Monte Porzio Catone, Italy.
        \and
        INAF-Osservatorio Astronomico di Bologna, via Ranzani 1, I-40127 Bologna, Italy.
        \and
        Institute for Computational Cosmology, Department of Physics, Durham University, South Road, Durham, DH1 3LE, England.
        \and
        Dipartimiento di Astronomia, Universit{\`a} di Padova, Vicolo dell'Osservatorio 3, 35122 Padova, Italy.
        \and        
        INAF - Osservatorio Astronomico di Trieste, via Tiepolo 11, 34143 Trieste, Italy
}

\date{Received ; accepted }

\abstract
{
Gravitational lensing by massive galaxy clusters allows study of
the population of intrinsically faint infrared galaxies that lie
below the sensitivity and confusion limits of current infrared and
submillimeter telescopes. 
We present ultra-deep PACS 100 and 160\,$\mu$m observations 
toward the cluster lens Abell 2218 to penetrate the
$Herschel$ confusion limit.
We derive source counts down to a flux density of 1\,mJy at 
100\,$\mu$m and 2\,mJy at 160\,$\mu$m, 
        aided by strong gravitational lensing.  At these levels,
        source densities are 20 and 10 beams/source in the two
        bands, approaching source density confusion at 160\,$\mu$m.
        The slope of the counts below the turnover of the
        Euclidean-normalized differential curve is constrained in
        both bands and is consistent with most of the recent backwards
        evolutionary models.
 By integrating number counts over the flux range accessed by 
   Abell 2218 lensing ($0.94-35$\,mJy at 100\,$\mu$m and 
   $1.47-35$\,mJy at 160\,$\mu$m), we retrieve a cosmic infrared background surface brightness 
   of $\sim$8.0 and $\sim$9.9 nW m$^{-2}$ sr$^{-1}$, in the respective
   bands. These values correspond to $55\pm24$\% and $77\pm31$\% 
   of DIRBE direct measurements. Combining Abell 2218 results 
   with wider/shallower fields, these figures increase 
   to $62\pm25$\% and $88\pm32$\% CIB total fractions, resolved at 100 and 
   160\,$\mu$m, disregarding the high uncertainties of DIRBE
    absolute values.
}

\keywords{Surveys -- Infrared: galaxies --  Galaxies: evolution -- Galaxies: high-redshift -- clusters: general -- Gravitational lensing: strong}

\maketitle
%
%________________________________________________________________

\section{Introduction}

The discovery of the cosmic infrared background (CIB) \citep{puget96,fixsen96,lagache99}
has opened new perspectives on galaxy formation and evolution.
A large number of the sources contributing to this far-infrared (FIR) CIB have been resolved in the mid-infrared (mid-IR) at 15\,$\mu$m with the ISOCAM instrument on ISO
 \citep{genzel00,elbaz02}, and later at 24$\mu$m with MIPS on {\it Spitzer} \citep{papovich04,dole04,frayer06}.
A striking result concerns the evolution of the infrared and submillimeter (sub-mm) galaxy population: 
the infrared source counts are higher than no-evolution or moderate-evolution models  
and provide strong constraints on the evolution of the bolometric energy output from distant galaxy populations.

However, at FIR and sub-mm wavelengths, a much lower fraction has been resolved so far, 
because of the small aperture of telescopes, the prohibitive confusion limits, and the low
sensitivity of available instruments. Nevertheless, stacking results
(Dole et al. 2006) show that most or all of the FIR background is
due to known high-z IR galaxies.

\begin{figure*}[t]
\centerline{ 
\includegraphics[width=0.5\textwidth]{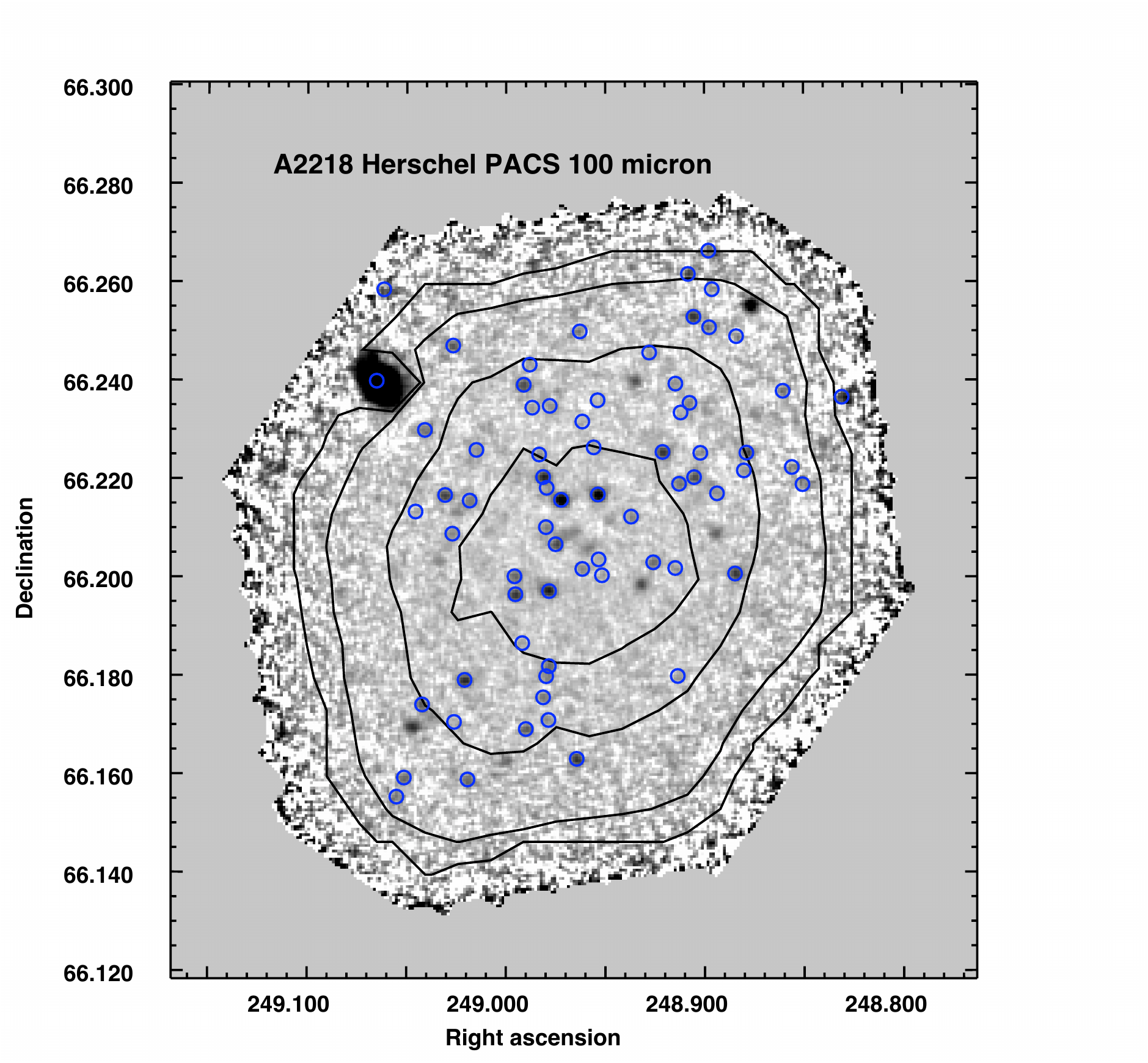}
\includegraphics[width=0.5\textwidth]{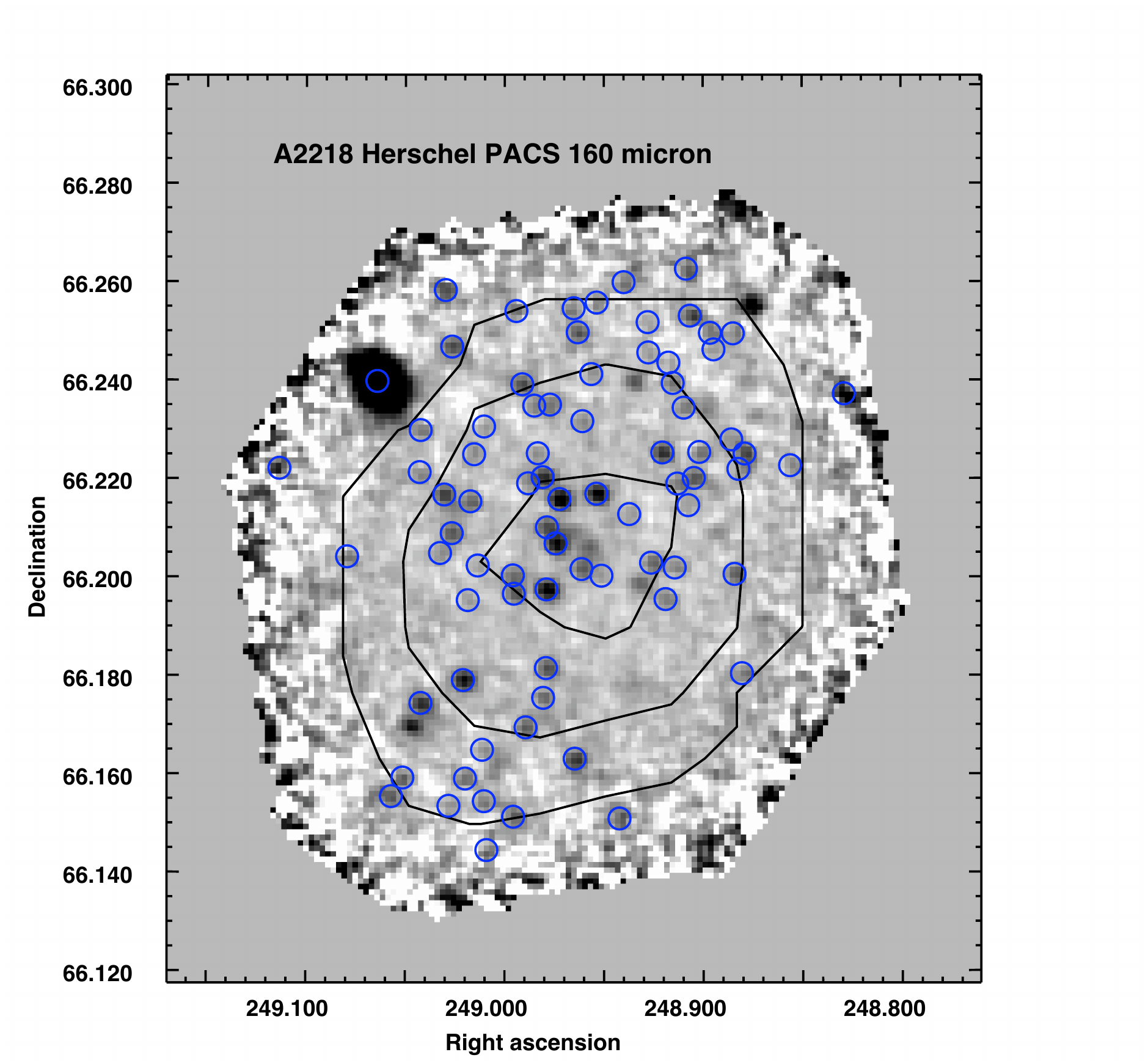}
}
  \caption{(LEFT) PACS 100\,$\mu$m map toward Abell 2218 with lensed
and foreground sources marked by blue circles. Other sources are
identified cluster members. The rms contours at 0.7, 1.0, 2.0, and 4.0\,mJy are overlaid in black. (RIGHT) PACS 160\,$\mu$m map. The rms contours at 1.4, 2.0, and 4.0\,mJy are overlaid in black.}
   \label{fig:a2218maps}
\end{figure*}

The {\it Herschel Space Observatory}$\textquoteright$s
\citep{pilbratt10} compact point-spread-function (PSF) enables probes of the FIR emission of
large samples of galaxies near their spectral energy distribution
(SED) peak, over a wide z range.  The sensitive {\it Photometer
Array Camera \& Spectrometer} (PACS) \citep{poglitsch10}  achieves
imaging surveys of unrivalled depth at 70, 100, and 160\,$\mu$m in the
PEP (PACS Extragalactic Probe) Key Program (PI D. Lutz).

As pioneered by \citet{smail97} in the sub-mm, and later in the
mid-IR \citep{altieri99,metcalfe03}, we are conducting a survey 
toward massive clusters as part of PEP, to resolve the FIR beyond the
field-survey limit and penetrate the $Herschel$ confusion limit with
powerful gravitational lenses. This letter reports initial
observations on the cluster Abell~2218.

\vspace*{-0.25cm}

\section{Observations}\label{sec:data}

The rich and massive lensing cluster Abell 2218 was targeted 
in the PEP science demonstration phase. This field has been intensively
studied, including deep archival Hubble Space Telescope\,(HST)/ACS and {\it Spitzer} (IRAC \& MIPS)
observations, which allow simple and direct identification of the
sources detected with $Herschel$/PACS.

Our 13 hours of observations, centered on the cD galaxy at
RA=16h35m51.84s, Dec=62d14m19.2s (J2000), used scan mapping
(20$^{\prime\prime}$/s scan speed; 4$^{\prime}$ scan leg length;
20$^{\prime\prime}$ cross-scan steps) over an area of about
6$^{\prime}$x6$^{\prime}$, with coverage strongly peaked toward the
center. 
Though affected by relatively high overheads (reduced later
in the mission), they are the deepest observations so far at 100 and
160\,$\mu$m.

\vspace*{-0.25cm}

\section{Data analysis}\label{sec:analysis}

The data were reduced using HIPE \citep{ott10} v2.0. build 1328. 
The 10\,Hz data cubes were processed with the standard PACS
pipeline, along with custom procedures: 1) to remove interference patterns,
transients from calibration blocks and tracking anomalies, 2) to perform re-centering of positional offsets.  Moreover, to remove detector
drifts and $1/f$ noise, a sliding high-pass filter was run on the
pixel timelines with an iterative masking of the brighter
sources.
For more information, refer to
\citet{berta10}. Final maps are displayed in Figure
\ref{fig:a2218maps}.

Source extraction was performed with the StarFinder PSF-fitting code \citep{diolaiti00}, down to the 3$\sigma$ level.
The total number of sources is reported in Table \ref{tab:tabsources}.
Based on random extractions from the residual images, the averaged 1$\sigma$ noise levels in the maps are 
0.92\,mJy at 100\,$\mu$m and 1.61\,mJy at 160\,$\mu$m.

\begin{table}[hb]
\caption{Properties of the A2218 PACS sources detection.}
\label{tab:tabsources}
\centering 
\begin{tabular} {llc}
\hline
               & 100\,$\mu$m & 160\,$\mu$m   \\
\hline
PSF FHWM (arcsec) & 7.5 & 11.0 \\
\hline
\hline
$1\sigma$ average sensitivity (mJy) & 0.92 & 1.61 \\
$1\sigma$ central sensitivity (mJy) & 0.6 & 1.2 \\
\hline
\hline
number of detected sources ($>3\sigma$)        &  98 & 94 \\
number of lensed sources ($>3\sigma$) &  70 & 78 \\
\hline
\hline
lensed sources with spectroscopic redshift & 9& 13   \\ 
\hline
lensed sources with photometric redshift & 20 &17   \\ 
\hline
\hline
\end{tabular}
\end{table}

We inspected all sources in the maps by eye.
The source catalog was cross-correlated with various redshift
surveys for Abell 2218
(\citealt{leborgne92,ziegler01,ebbels98,metcalfe03}; Kneib, priv.
comm.) and with photometric redshifts of the Abell~2218 MIPS
24$\mu$m sources computed using HST/ACS images, near-IR images and the 4
{\it Spitzer}/IRAC bands \citep{ciesla09}. All PACS sources were detected
in the {\it Spitzer}/MIPS observation. This cross-correlation allowed us
to classify sources as foreground, cluster, or background galaxies. 
More than 90\% of background sources with confirmed spectroscopic or photometric redshift 
have $F_{160}/F_{100} \gtrsim 1$, while 11 out of the 13 confirmed cluster galaxies detected both at 100 and 160\,$\mu$m have  $F_{160}/F_{100} < 1$.
This allows classification of sources with $F_{160}/F_{100} \gtrsim 1$, but without redshift, as lensed-background galaxies. This is as also supported by their extremely faint magnitudes in the optical and/or their disturbed morphologies.

Less than 30\% (20\%) of the sources at 100\,$\mu$m  (160\,$\mu$m) are classified as cluster galaxies.
This situation is very similar to the mid-IR, 
where e.g. \citet{altieri99, metcalfe03, egami06,hopwood10}, 
showed that the large majority of far-IR sources are background, hence lensed, sources. 
The cluster core is virtually transparent in the FIR and acts as a natural telescope to provide
a magnified view of the background sky.
This increases the sensitivity and at the same time reduces the effects of source confusion;
in particular, the cD galaxy is undetected in both bands.

Four of the 7 sub-mm sources in the ultra-deep map of Abell
2218 \citep{knudsen06} are detected in the 160\,$\mu$m map, including
the triple z=2.516 sub-mm source SMMJ16359+6612 \citep{kneib04a},
the sources SMMJ163555.2+661150 at $z=1.034$, and
SMMJ163541.2+661144 at $z=3.1824$ \citep{knudsen09}. However, the highest
redshift sub-mm source (z=4.048) SMMJ163555.5+661300 and the two
faintest sub-mm sources have not been detected, which is also true for the
z=5.56 source of \citet{ellis01} and the $z\sim7$ triple source
of \citet{kneib04b}.

\vspace*{-0.25cm}

\section{Lensing inversion and source counts}\label{sec:counts}

\begin{figure}[b]
\includegraphics[width=0.47\textwidth]{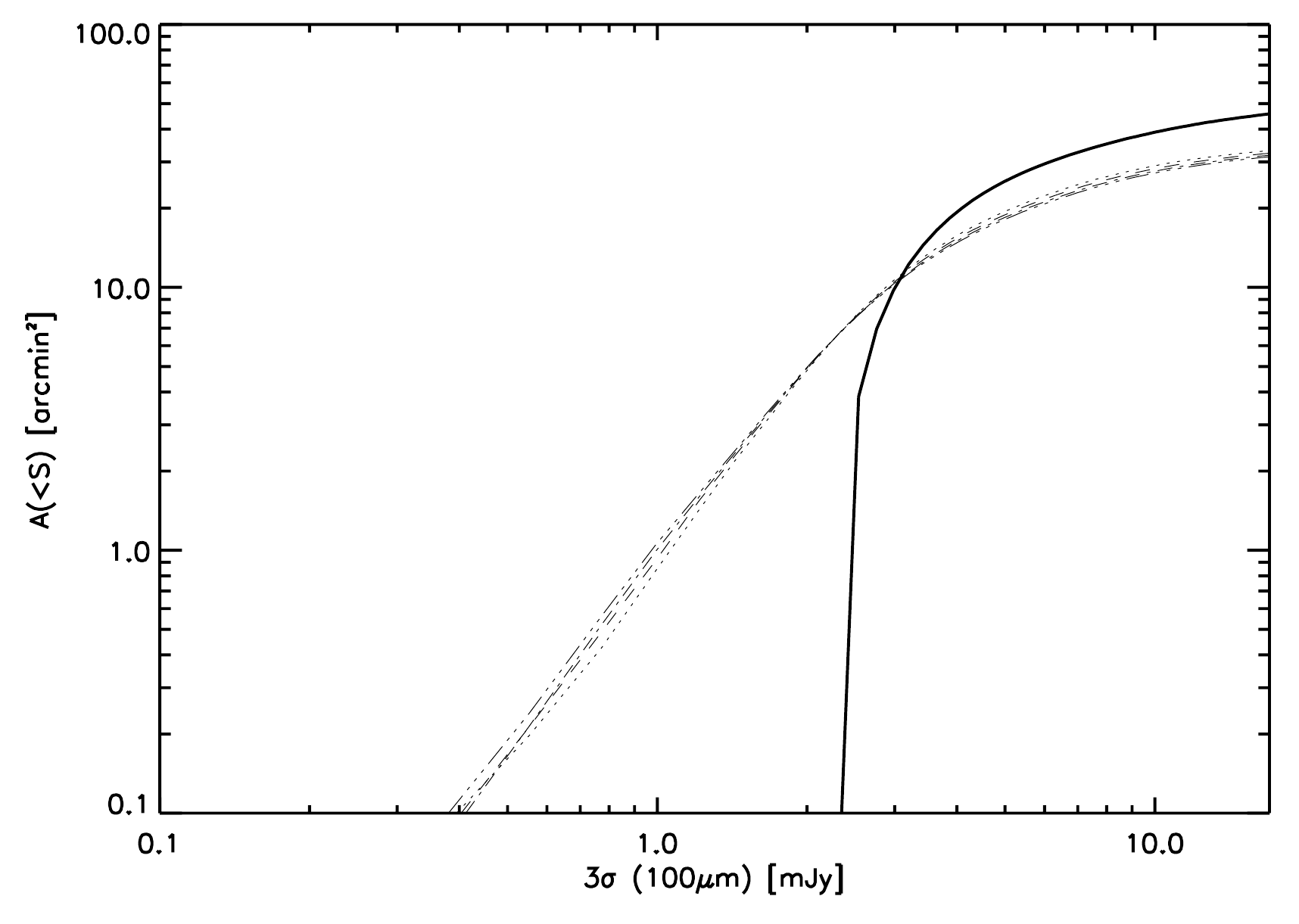}

  \caption{The area of the source plane at 100$\mu$m, as a function of the 3$\sigma$ sensitivity, for redshifts z=0.7, 1, 1.5,  and 2.5, 
compared to the corresponding area in the image plane (i.e. apparent sky - the solid line).}
  \label{fig:areasens}
\end{figure}

Lensing acts in two ways on the background sky:
{\it i)} it amplifies source brightness,
typically by a factor of 2, but by as much as 10 near critical lines;
{\it ii)} it magnifies the area probed - 
and both the flux amplification and the space magnification are stronger
toward the cluster core and increase with source-plane redshift.
We exploited the detailed mass model of Abell 2218
\citep{kneib96,kneib04b}, which considers 8 multiply imaged
systems, of which 7 have spectroscopically confirmed redshifts (among these
the several high-z multiply-lensed sources mentioned at the end of Section 3).

We used the publicly available software, LensTool, described in \citet{jullo07}
to compute the source amplification factors and the surface magnification due to lensing.
Where redshifts were not available for sources judged by morphology
or intrinsic faintness  to be in the background, a value of $z=0.7$
was assigned, corresponding to  the median redshift of field sources
having spectroscopic or photometric redshift. This has little impact
on the lensing amplification factor and area magnification,  as
illustrated in Figure \ref{fig:areasens}. This figure shows the area
mapped to a given sensitivity in the image plane (i.e. on the apparent sky), 
compared to the corresponding areas in source planes at redshifts 0.7, 1, 1.5, and 2.5. 
For redshifts above 0.7, the lensing surface magnification depends only weakly on
redshift. The lensing amplification factor for a given source on the
other hand may be more sensitive to the redshift. However, setting all
(morphologically classified) background sources without known redshift to z=1 
(instead of z=0.7) did not change the shape of the counts significantly, because it
shifted only a few sources among adjacent flux bins.

Figure \ref{fig:areasens} expresses the fact that some small, highly-lensed regions of the 
source plane are mapped onto larger areas in the image plane (the apparent sky), with the result
that, over a few arcmin$^{2}$, flux densities in the range 1 to 3\,mJy become accessible by virtue of 
the lensing effect.

Ten sources (12 apparent sources, including all images of SMM J16359+6612) have lens-corrected fluxes below 3\,mJy at 100\,$\mu$m and 13 (15 apparent)  below 5.7\,mJy at 160\,$\mu$m, 
with these quoted limits the 3$\sigma$ sensitivities achieved on the GOODS-N field \citep{berta10}.
Such sources would most likely not be detected even in upcoming deeper blank-field surveys, 
like GOODS-S (deeper by a factor of 2).
The triple source SMM\,J16359+6612, for instance, is amplified by a factor of 45, in total. 
It was counted only once, as number counts refer to the source plane.

By correcting for lensing amplification, surface magnification
effects, contamination by cluster galaxies, and non-uniform
sensitivity of our maps, we can derive number counts at 100\,$\mu$m
and 160\,$\mu$m. Because of the non-uniform sensitivity of the
maps on the sky and the lensing effect, different areas on
the sky are surveyed to different depths in the source plane. The
object density per flux bin was computed using gain-dependent 
surface areas. Incompleteness affects the counts progressively 
below apparent fluxes of 6\,mJy at 100\,$\mu$m and 9\,mJy at 160\,$\mu$m.
We restricted these counts to 4$\sigma$ detections where the
completeness  of measurement is typically 80\%.  This avoids
potentially complicated folding of completeness correction with
lensing correction, not justified by the relatively small numbers of
sources and relatively large statistical error bars associated with
these sample counts on a single lensing cluster.
A comprehensive
completeness analysis will be required when combining the
observations of a sample of massive cluster lenses to extract the
galaxy number counts below the 1\,mJy level.

\begin{figure*}[t]
\centerline{
\includegraphics[width=0.45\textwidth]{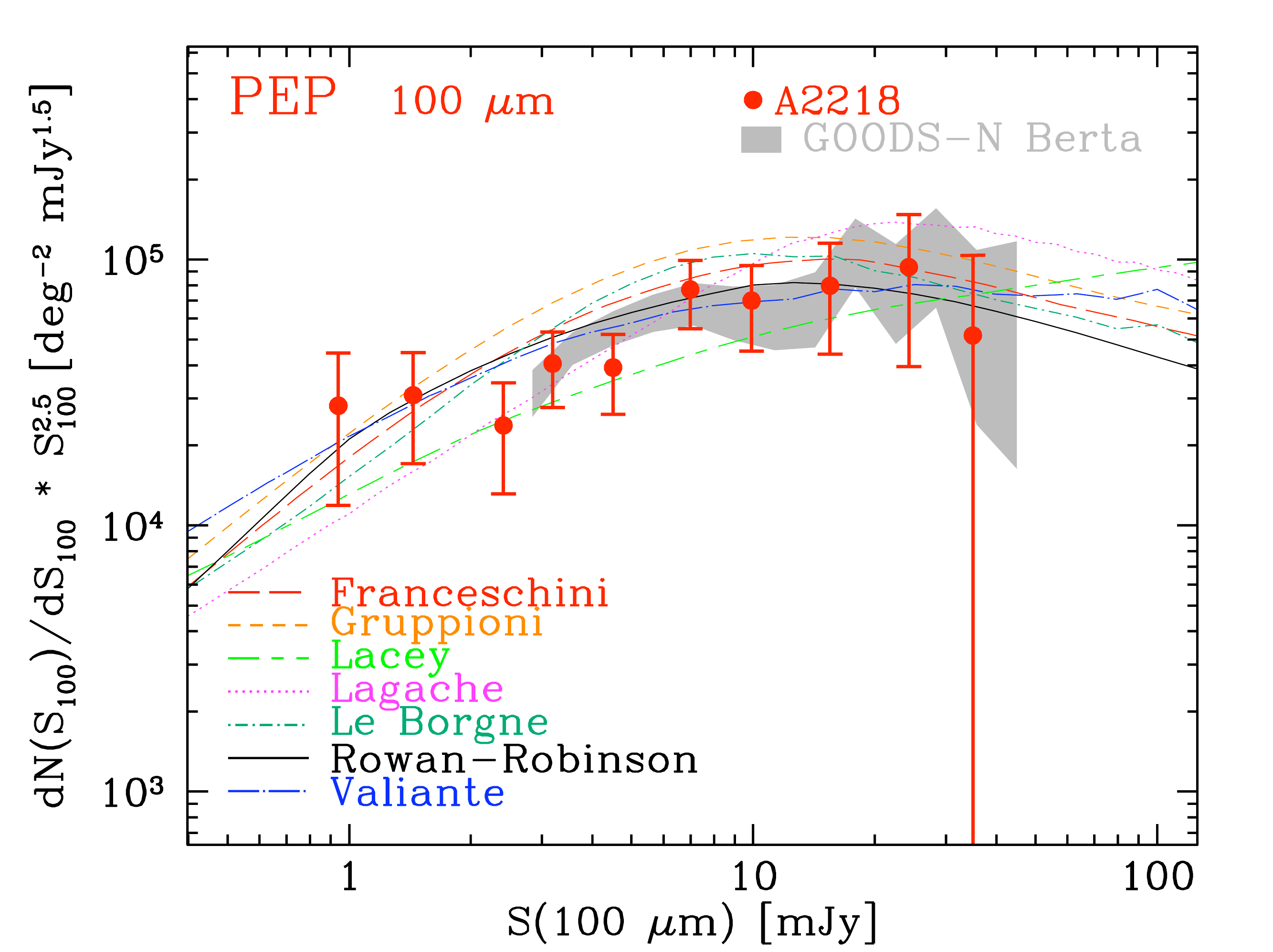}
\includegraphics[width=0.45\textwidth]{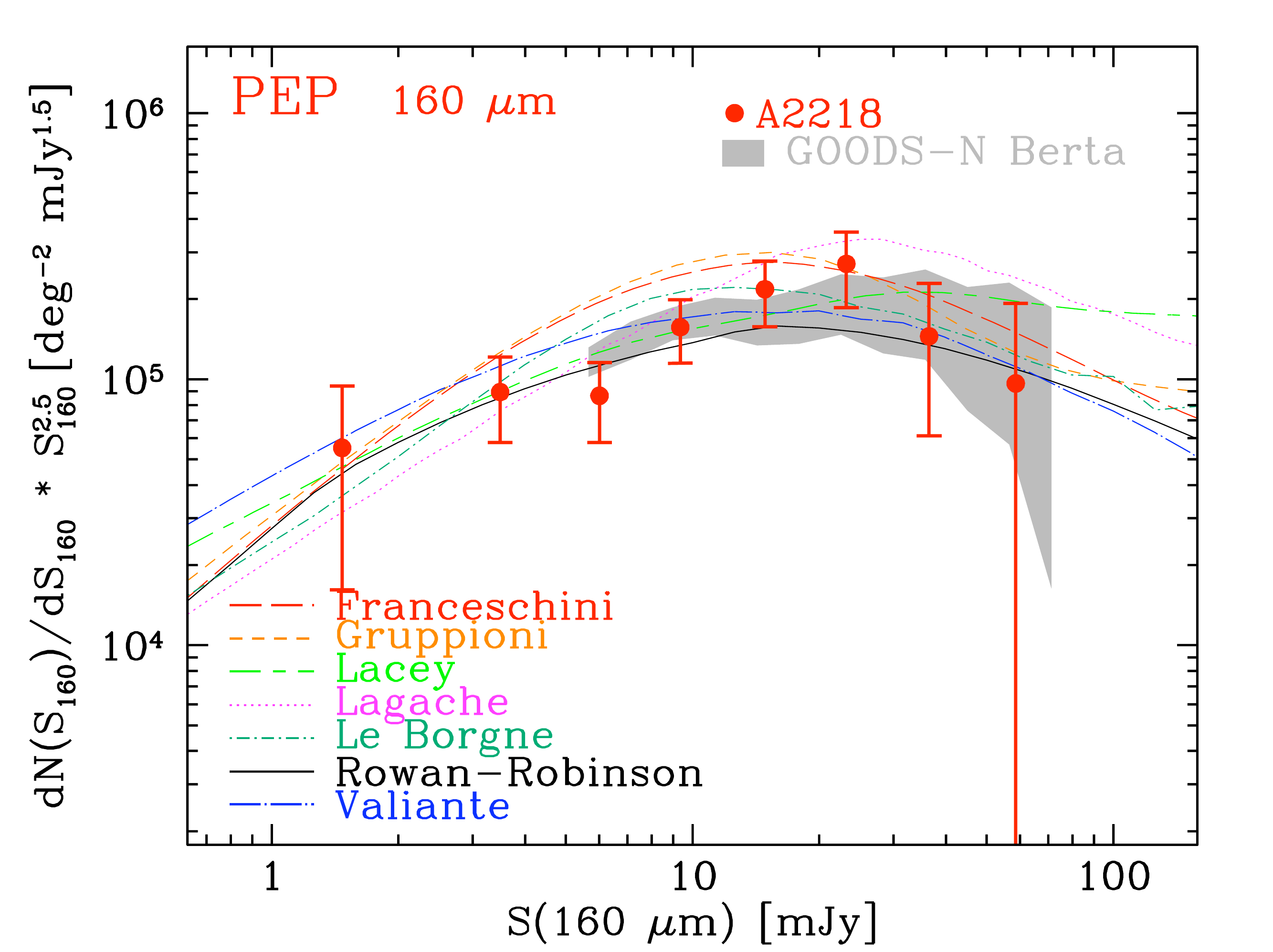}
}

  \caption{Number counts at 100 and 160\,$\mu$m with lensing
  correction (red filled
  circles), normalized to the Euclidean slope, against the prediction of various synthetic counts models. 
  See references  in  \citet{berta10}. 
  Errors refer to pure Poisson statistics at 68\% confidence limit GOODS-N counts from Berta et al. are contrasted in
  the shaded area. 
  }
  \label{fig:counts100}
\end{figure*}

Our counts are barely affected by confusion at 100\,$\mu$m, with a source density
of 20 beams/source in the image plane (for the \citealt{lagache03} definition of the beam).
At 160\,$\mu$m
the density is as high as 10 beams/source, with even fewer beams per source 
in the central area of high-lensing. Hence, the Abell 2218 catalog is affected by source 
confusion at 160\,$\mu$m 
(16.7 beams/source from \citet[][]{dole03}) and the high density of detected sources 
prevents the extraction of fainter objects.

The source counts, corrected for cluster contamination and lensing effects, in both the 100\,$\mu$m and 160\,$\mu$m bands,
are presented in Figure \ref{fig:counts100}, normalized to the Euclidean slope ($dN/dS \propto S^{-2.5}$).  
Error bars consider Poisson statistics only, as flux uncertainties are minimal by comparison.

\vspace*{-0.25cm}

\section{Discussion}\label{sec:discussion}

The Abell 2218 $Herschel$/PACS maps are the deepest FIR maps to date.
Covering an area of $6\times 6$ arcmin$^{2}$ over the cluster core,
we have surveyed the region where strongly lensed background sources are present.
At fluxes higher than 5\,mJy, the lens-corrected normalized number counts are consistent, within error bars, with the GOODS-N 
counts \citep{berta10} showing evidence of a peak between 5-10 and 30\,mJy.
Thanks to gravitational lensing, we could extend the counts down 
to $\sim$1\,mJy at 100\,$\mu$m and $\sim$2\,mJy at 160\,$\mu$m, confirming the negative 
sub-Euclidean slope below the turnover.

The Abell 2218 differential counts (not normalized to the Euclidean slope) at 100\,$\mu$m show a faint-end slope similar to GOODS-N, but they are higher at fainter fluxes.
At 160\,$\mu$m the differential counts show a steeper slope than GOODS-N ($-\,1.82$ instead of $-1.67$ assuming the functional form of the counts: $dN/dS \propto S^\alpha$).
This is also reflected in the Euclidean-normalized counts. The high error bars seen 
at the faint end come from the poor statistics.
The counts could be slightly underestimated  at 160\,$\mu$m at the lowest fluxes because 
the source density is approaching confusion with no correction made for this.

The counts are reproduced well by the models of \citet{valiante09} or  \citet{rowanrobinson09}, both at 100 and 160\,$\mu$m,
but other backwards evolutionary models similarly reproduce the downward turn below 10\,mJy.
The summed contribution of resolved galaxies provides a lower limit to the IR background and can be compared to the
estimation of the CIB. Here we have adopted the latest  measure of its surface brightness from COBE/DIRBE maps \citep{dole06}: 
$14.4\pm6.3$ nW\,m$^{-2}$\,sr$^{-1}$
at 100\,$\mu$m and 
$12.75\pm6.4$ nW\,m$^{-2}$\,sr$^{-1}$
 at 160\,$\mu$m (which is an interpolated value).

\begin{table}
\caption{Resolved CIB surface brightness for Abell 2218,
and combined with the COSMOS contribution in the higher flux range ("PEP").
}
\label{tab:tabCIB}
\centering 
\begin{tabular} {|l c c c c|}
\hline
		 		&  flux-range 	&   CIB 	& d(CIB)	&	fraction   \\
				   &  (mJy)  & (nW m$^{-2}$ sr$^{-1}$) & & (\%) \\
\hline
100\,$\mu$m A2218			&  0.94-35	&  7.9	& 3.4	&     55 $\pm$ 24  \\
100\,$\mu$m PEP			&  0.94-142	& 8.8 	& 3.5	&     62 $\pm$ 25  \\
\hline
160\,$\mu$m A2218			& 1.47-36		&  9.9 		& 3.9	&     77 $\pm$ 31 \\
160\,$\mu$m PEP			&  1.47-179	& 11.3		& 4.1	&     88 $\pm$ 32  \\
\hline
\end{tabular}
\end{table}

The contribution to the CIB by Abell 2218 background galaxies above
    the 4$\sigma$ detection threshold was computed by simply integrating the observed number
    counts (Table\,\ref{tab:tabCIB}). Errors were computed by
    integrating the envelope of the counts with their uncertainties. More than
    half of the DIRBE CIB \citep{dole06} has been directly 
    resolved, which is also consistent with the surface
    brightness found by \citet{bethermin10} in a stacking analysis of {\it Spitzer}/MIPS sources.

    Combining the deep counts in Abell 2218 with the results obtained by 
    Berta et al (2010) in wider/shallower PEP fields (e.g. COSMOS), we were 
    able to extend the integration flux range to 142\,mJy at 100\,$\mu$m and 179\,mJy at  
    160\,$\mu$m. Consequently, the resolved CIB fractions 
    increase to $62\pm25$\% and $88\pm32$\% in the two bands. 
    One must keep in mind that not only is the CIB surface brightness 
    from PEP affected by large uncertainties (cosmic variance), but the 
    reference values by \citet{dole06} are also defined only within 
    a factor of $\sim$2.

Another 9 massive lensing clusters will be targeted as part of PEP and another 40 more 
in the open-time $Herschel$ Lensing Survey KP
\citep{egami10}. We expect that, in the coming years, combined results from many lensing clusters will greatly improve the statistics of highly amplified sources and constrain source densities around $\sim$1\,mJy or below. By penetrating below the unlensed $Herschel$ confusion limit and probing the high-redshift galaxy populations beyond the sensitivity limit of blank-field surveys, the fraction of the resolved CIB will increase.

\begin{acknowledgements}
PACS has been developed by a consortium of institutes led by MPE (Germany) and 
including UVIE (Austria); KU Leuven, CSL, IMEC (Belgium); CEA, LAM (France); 
MPIA (Germany); INAF-IFSI/OAA/OAP/OAT, LENS, SISSA (Italy); IAC (Spain). 
This development has been supported by the funding agencies BMVIT (Austria), 
ESA-PRODEX (Belgium), CEA/CNES (France), DLR (Germany), ASI/INAF (Italy), 
and CICYT/MCYT (Spain).

\end{acknowledgements}

\end{document}